\documentclass[
    ,final            
  ]
  {aipproc}

\layoutstyle{6x9}

\usepackage{graphicx}

\begin{document}

\title{Charm meson resonances in $D$ semileptonic decays}



\author{Svjetlana Fajfer}{address={J. Stefan Institute, Jamova 39, P. O. Box 3000, 1001 Ljubljana, Slovenia},altaddress={Department of Physics, University of Ljubljana, Jadranska 19, 
1000 Ljubljana, Slovenia},email={svjetlana.fajfer@ijs.si}}

\author{Jernej Kamenik}{address={J. Stefan Institute, Jamova 39, P. O. Box 3000, 1001 Ljubljana, Slovenia},email={jernej.kamenik@ijs.si}}

\begin{abstract}
Motivated by recent experimental results we reconsider semileptonic $D
\to P \ell \nu_{\ell} $ and $D\to V \ell \nu_{\ell} $ decays within a model which combines heavy quark
symmetry and properties of the chiral Lagrangian. Using limits of 
soft collinear effective theory and heavy quark effective 
theory we parametrize the semileptonic form factors. We include excited
charm meson states in our Lagrangians and determine their
impact on the charm meson semileptonic form factors. 
Then we calculate branching ratios for all 
$D \to P \ell \nu_{\ell} $ and $D \to V l \nu_l$ decays.
\end{abstract}

\maketitle


The knowledge of the form factors which describe the weak $heavy \to
light$ semileptonic transitions is very important for the accurate
determination of the CKM parameters from the experimentally measured
exclusive decay rates. Usually, the attention has been devoted to 
$B$ decays and the determination of the phase of the $V_{ub}$ CKM matrix element.
At the same time in the charm sector, the most accurate determination of the size of $V_{cs}$ and $V_{cd}$ matrix elements is not from a direct measurement, mainly due to theoretical uncertainties  in the calculations of the relevant form factors' shapes.
\par 
Recently, there have been new interesting results on $D$-meson
semileptonic decays.  The CLEO and FOCUS collaborations have studied semileptonic
decays $D^0\rightarrow \pi^- \ell^+ \nu$ and $D^0\rightarrow K^- \ell^+ \nu$~\cite{Huang:2004fr,Link:2004dh}. Their data provide new information on the $D^0\rightarrow \pi^- \ell^+ \nu$ and $D^0\rightarrow K^- \ell^+ \nu$ form factors. Usually in $D$ semileptonic decays a simple pole parametrization was used in the past. The results of Refs.~\cite{Huang:2004fr,Link:2004dh} for the single pole parameters required by the fit of their data, however, suggest pole masses, which are inconsistent with the physical masses of the lowest lying charm meson resonances. In their anlyses they also utilized a modified pole fit as suggested in~\cite{Becirevic:1999kt} and their results indeed suggest the existence of
contributions beyond the lowest lying charm meson resonances~\cite{Huang:2004fr}.
\par
In addition to these results new experimental studies of 
charm meson resonances have provided a lot of new information on the charm sector~\cite{Aubert:2003fg,Vaandering:2004ix,Besson:2003jp,Evdokimov:2004iy} which we can now apply to $D$ and $D_s$ semileptonic decays.
\par
The purpose of our studies~\cite{Fajfer:2004mv, Fajfer:2005ug} is to accommodate  
contributions of the newly
discovered and theoretically predicted charm mesons in form factors which 
are parametrized
using constraints coming from heavy quark effective theory (HQET) limit for the region of 
$q_{max}^2$ and in the $q^2 \simeq 0$ region using results of 
soft collinear effective theory (SCET).
We restrain our discussion to the leading chiral and $1/m_H$ terms in the expansion.

\par

The standard decomposition of the current matrix elements 
relevant to semileptonic decays between a heavy pseudoscalar meson state 
$|H(p_H)\rangle$ with momentum $p_H^{\nu}$ and a light pseudoscalar meson state $| P (p_P) \rangle$ with momentum $p_P^{\mu}$ is in terms of two scalar functions of the exchanged momentum squared $q^2 = (p_H-p_P)^2$ -- the form factors $F_+(q^2)$ and $F_0(q^2)$. Here $F_+$ denotes the vector form factor and it is dominated by vector meson resonances, while $F_0$ denotes
the scalar form factor and is expected to be dominated by scalar meson
resonance exchange~\cite{Marshak:1969tw,Wirbel:1985ji}. 
In order that the matrix elements are finite at $q^2=0$, the form factors must
also satisfy the relation $F_+(0)=F_0(0)$.
\par
The transition of $|H(p_H)\rangle$ to light vector meson 
$|V(p_V,\epsilon_V)\rangle$ with momentum 
$p_V^{\nu}$ and polarization vector $\epsilon_V^{\nu}$ is similarly parameterized in terms of four form factors $V$, $A_0$, $A_1$ and $A_2$, again functions of the exchanged momentum squared $q^2 = (p_H-p_V)^2$. Here $V$ denotes the vector form factor and is expected to be dominated by vector meson resonance exchange, the axial $A_1$ and $A_2$ form factors are expected to be dominated by axial resonances, while $A_0$ denotes the 
pseudoscalar form factor and is expected to be dominated by pseudoscalar 
meson resonance exchange~\cite{Wirbel:1985ji}.
As in previous case in order that the matrix elements are finite at $q^2=0$, the form factors must also satisfy the well known relation $A_0(0)+A_1(0)(m_H+m_V)/2m_V-A_2(0)(m_H-m_V)/2m_V=0$.
\par
Next we follow the analysis of Ref.~\cite{Becirevic:1999kt}, where the $F_+$
form factor in $H\to P$ transitions is given as a sum of two pole contributions, while the
$F_0$ form factor is written as a single pole. This 
parametrization includes all known properties of form factors at large $m_H$.  Using a relation which connects the  form factors within large energy release approach~\cite{Charles:1998dr} the authors in Ref.~\cite{Becirevic:1999kt} propose the following form factor 
parametrization
\begin{equation}
F_+(q^2)=\frac{c_H(1-a)}{(1-x)(1-a x)}, \qquad F_0(q^2)=\frac{c_H(1-a)}{1-b x},\label{f_+_dipole}
\end{equation}
where $x=q^2/m_{H^*}^2$. 
\par
Utilizing the same approach we propose a general parametrization of the heavy to light vector form factors, which also takes into account all the known scaling and resonance 
properties of the form factors.
As already mentioned, there exist the well known HQET scaling laws in 
the limit of zero recoil~\cite{Isgur:1990kf} while in the SCET limit $q^2\to 0$  one obtains that all four $H \to V$ form factors can be related to only two universal SCET 
scaling functions~\cite{Charles:1998dr}.
\par
The starting point is the vector form factor $V$, which is dominated by the pole at $t=m_{H^*}^2$ when considering the part of the phase space that is close to the zero recoil. For the $heavy\to light$ transitions this situation is expected to be 
realized near the zero recoil where also the HQET scaling  applies. 
On the other hand, in the region of large recoils, SCET dictates the 
scaling described in \cite{Charles:1998dr}. In the full analogy with the 
discussion made in Refs. \cite{Becirevic:1999kt, Hill:2005ju}, the vector 
form factor consequently receives contributions from two poles and can be 
written as
\begin{equation}
V(q^2) = \frac{c'_H (1-a)}{(1-x)(1-a x)},
\label{eq_v_ff}
\end{equation}
where $x=q^2/m_{H^*}^2$ ensures, that the form factor is 
dominated by the physical $H^*$ pole, while $a$ measures 
the contribution of higher states which are parametrized by another 
effective pole at $m_{\mathrm{eff}}^2=m_{H^*}^2/a$.  
\par
An interesting and useful feature one gets from the 
SCET is the relation between $V$ and $A_1$~\cite{Charles:1998dr,Ebert:2001pc,Burdman:2000ku, Hill:2004rx} at $q^2\approx 0$. When combined with our result~(\ref{eq_v_ff}), it imposes a single pole structure on $A_1$. We can thus continue in the same line of argument and write
\begin{equation}
A_1(q^2) =  \xi \frac{c'_H(1-a)}{1-b' x}.
\label{eq_a1_ff}
\end{equation}
Here $\xi=m_H^2/(m_H+m_V)^2$ is the proportionality factor between $A_1$ and $V$ from the SCET relation, while $b'$ measures the contribution of resonant states with spin-parity assignment $1^+$ which are parametrized by the effective pole at $m_{H'^*_{\mathrm{eff}}}^2=m_{H^*}^2/b'$. It can be readily checked that also $A_1$, when parametrized in this way, satisfies all the scaling constraints. 
\par
Next we parametrize the $A_0$ form factor, which is completely 
independent of all the others so far as it is dominated by the pseudoscalar 
pole and is proportional to a different universal function in SCET. 
To satisfy both HQET and SCET scaling laws we parametrize it as 
\begin{equation}
A_0(q^2) = \frac{c''_H (1-a')}{(1-y)(1-a' y)},
\label{eq_a0_ff}
\end{equation}
where $y = q^2/m_H^2$ ensures the physical $0^-$ pole dominance at small 
recoils and $a'$ again parametrizes the contribution of 
higher pseudoscalar states by an effective pole at 
$m_{H'_{\mathrm{eff}}}^2=m_{H}^2/a'$. The resemblance to $V$ is 
obvious and due to the same kind of analysis~\cite{Becirevic:1999kt} although 
the parameters appearing in the two form factors are completely unrelated. 
\par
Finally for the $A_2$ form factor, due to the pole behavior of the $A_1$ 
form factor on one hand and different HQET scaling at 
$q^2_{\mathrm{max}}$ on the other hand, we have to go beyond 
a simple pole formulation. Thus we impose
\begin{equation}
A_2(q^2) = \frac{(m_H+m_V) \xi c'_H (1-a) + 2 m_V c''_H (1-a')}{(m_H-m_V)(1-b' x)(1-b'' x)},
\label{eq_a2_ff}
\end{equation}
which again satisfies all constraints. Due to the relations between the form factors we only gain one parameter in this formulation, $b''$. This however causes the contribution of the $1^+$ resonances to be shared between the two effective poles in this form factor.
\par
At the end we have parametrized the four $H\to V$ vector form factors in 
terms of the six parameters $c'_H$, $a$, $a'$, $b'$, $c''_H$ and 
$b''$.

\par

In our heavy meson chiral theory (HM$\chi$T) calculations we use the leading order heavy meson chiral  Lagrangian in which we include additional charm meson resonances.
The details of this framework are given in \cite{Fajfer:2004mv}  and \cite{Fajfer:2005ug}. 
We first calculate values of the form factors in the small recoil region. 
The presence of charm meson resonances in our Lagrangian affects the values of
the form factors at $q^2_{\mathrm{max}}$ and induces saturation of the second 
poles in the parameterizations of the $F_+(q^2)$, $V(q^2)$ and $A_0(q^2)$ form 
factors by the next radial excitations of $D_{(s)}^*$ and $D_{(s)}$ mesons respectively. 
Although the $D$ mesons mat not be considered heavy enough, we employ these parameterizations with model matching conditions at $q^2_{\mathrm{max}}$.
Using HQET parameterization of the current matrix elements~\cite{Fajfer:2004mv,Fajfer:2005ug}, which is especially suitable for HM$\chi$T calculations of the form factors near zero recoil, we are able to extract consistently the contributions of individual resonances from our Lagrangian to the various $D\to P$ and $D\to V$ form factors. 
We use physical pole masses of excited state charmed mesons in the extrapolation, giving for 
the pole parameters $a=m_{H^{*}}^2/m_{H'^{*}}^2$, $a'=m_{H}^2/m_{H'}^2$, $b'=m_{H^*}^{2}/m_{H_{A}}^2$.
Although in the general parameterization of the form factors the extra poles in 
$F_+$, $V$ and $A_{0,1,2}$ parametrized all the neglected higher resonances beyond the ground state heavy meson spin doublets $(0^-,1^-)$, we are here 
saturating those by a single nearest resonance.
The single pole $q^2$ behavior of the $A_1(q^2)$ form factor is explained 
by the presence of a single $1^+$ state relevant to each decay, while in 
$A_2(q^2)$ in addition to these states one might also account for their next 
radial excitations. However, due to the lack of data on their presence we 
assume their masses being much higher than the first $1^+$ states and we 
neglect their effects, setting effectively $b''=0$.
\par
The values of the new model parameters appearing in $D \to P l \nu_l$ 
decay amplitudes~\cite{Fajfer:2004mv} 
are determined by fitting the model predictions to known experimental
values of branching ratios $\mathcal B (D^0\rightarrow K^- \ell^+
\nu)$, $\mathcal B (D^+\rightarrow \bar K^0 \ell^+ \nu)$, $\mathcal B
(D^0\rightarrow \pi^- \ell^+ \nu)$, $\mathcal B (D^+\rightarrow \pi^0
\ell^+ \nu)$, $\mathcal B (D^+_s\rightarrow \eta \ell^+ \nu)$ and
$\mathcal B (D^+_s\rightarrow \eta' \ell^+
\nu)$~\cite{Eidelman:2004wy}. In our  calculations of decay widths we 
neglect the lepton mass, so the form factor $F_0$, which is
proportional to $q^{\mu}$, does not contribute. For the decay width we
then use the integral formula proposed in~\cite{Bajc:1995km} with the flavor mixing parametrization of the weak current defined in~\cite{Fajfer:2004mv}.
\par
Similarly in the case of $D \to V l \nu_l$ transitions we have to 
fix additional model parameters~\cite{Fajfer:2005ug} and we again 
use known experimental values of branching
 ratios $\mathcal B (D_0\rightarrow K^{*-}\ell^+\nu)$, $\mathcal 
B (D_s^+ \rightarrow \Phi\ell^+\nu)$, $\mathcal B (D^+\rightarrow 
\rho^0\ell^+\nu)$, $\mathcal B (D^+\rightarrow K^{*0}\ell^+\nu)$, as well 
as partial decay width ratios $\Gamma_L/\Gamma_T (D^+\rightarrow K^{*0}
\ell^+\nu)$ and $\Gamma_+/\Gamma_- (D^+\rightarrow K^{*0}\ell^+\nu)$
~\cite{Eidelman:2004wy}. We calculate the decay rates for polarized final light vector mesons using helicity 
amplitudes $H_{+,-,0}$ as in for example~\cite{Ball:1991bs}. By neglecting the lepton masses we again arrive at the integral expressions from~\cite{Bajc:1995km} with the flavor mixing parametrization of the weak current defined in~\cite{Fajfer:2005ug}.

\par

\begin{figure}[!b]
\begin{tabular}{cc}
\scalebox{0.7}{\includegraphics{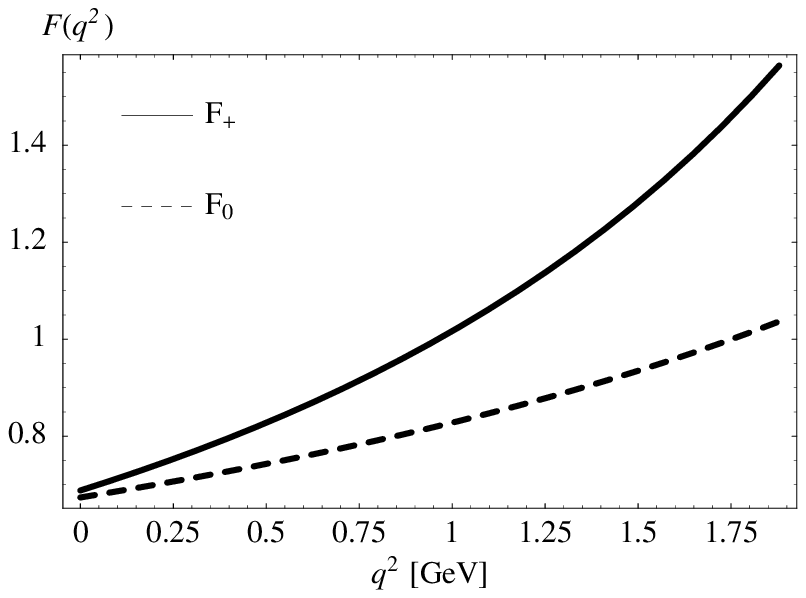}} & 
\scalebox{0.7}{\includegraphics{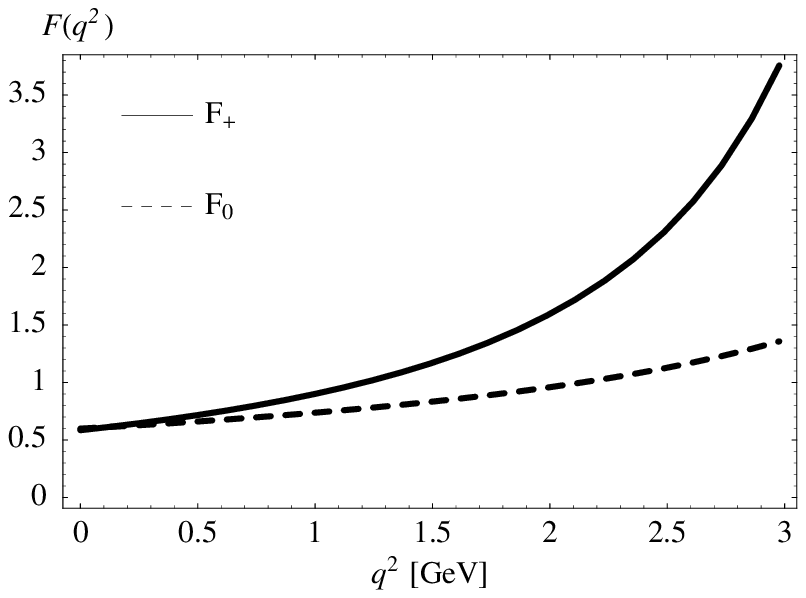}} \\
\multicolumn{2}{c}{\scalebox{0.7}{\includegraphics{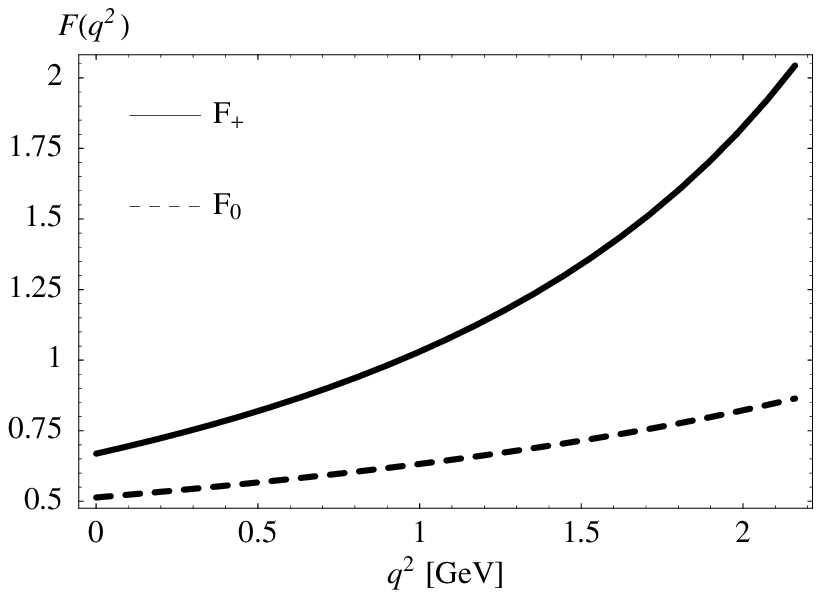}}} 
\end{tabular}
\caption{\label{FplotDK}$q^2$ dependence of the $D^0 \rightarrow K^-$ (upper left), $D^0 \rightarrow \pi^-$ (upper right) and $D_s \rightarrow K^0$ (lower) transition form factors.}
\end{figure}

We first draw  the $q^2$ dependence of the  $F_+$ and $F_0$ form factors for the 
$D^0\rightarrow K^-$, $D^0\rightarrow \pi^-$ and $D_s\rightarrow K^0$ transitions. The results are depicted in Fig.~\ref{FplotDK}.
Our model results, when extrapolated with the double pole parameterization, agree well with previous theoretical~\cite{Melikhov:2000yu,Aubin:2004ej} and experimental~\cite{Huang:2004fr,Link:2004dh} studies whereas the single pole extrapolation does not give satisfactory results.
Note that without the scalar resonance, one only gets a soft pion contribution to the $F_0$ form factor. This gives for the $q^2$ dependence of $F_0$ a constant value for 
all transitions, which largely disagrees 
with lattice QCD results~\cite{Aubin:2004ej} as well as heavily violates 
known form factor relations.
\par
We also calculate the
branching ratios for all the relevant $D\rightarrow P$ semileptonic
decays and compare the predictions of our model with experimental data
from PDG. The results are summarized in Table~\ref{PP_results_table}. 
For comparison we also include the results for the rates obtained with
our approach for $F_+(q_{\mathrm{max}}^2)$  but using a 
single pole fit. 
\begin{table}
\caption{\label{PP_results_table} The branching ratios for the $D\rightarrow P$ semileptonic decays. Comparison of our model fit with experiment as explained in the text.}
\begin{tabular}{l|ccc}
\hline
\tablehead{1}{c}{b}{Decay} & \tablehead{1}{c}{b}{$\mathcal{B}$ (model, double pole) [\%]} & \tablehead{1}{c}{b}{$\mathcal{B}$ (model, single pole) [\%]} & \tablehead{1}{c}{b}{$\mathcal{B}$ (Exp.~\cite{Eidelman:2004wy}) [\%]} \\
\hline 	
	$D^0\to K^-$ & $3.4$ & $4.9$ & $3.43 \pm 0.14$ \\
	$D^0\to \pi^-$ & $0.27$ & $0.56$ & $0.36 \pm 0.06$ \\
	$D_s^+\to \eta$ & $1.7$ & $2.5$ & $2.5 \pm 0.7$ \\
	$D_s^+\to \eta'$ & $0.61$ & $0.74$ & $0.89 \pm 0.33$ \\
	$D^+\to \bar K^0$ & $9.4$ & $12.4$ & $6.8 \pm 0.8$ \\
	$D^+\to \pi^0$ & $0.33$ & $0.70$ & $0.31 \pm 0.15$ \\
	$D^+\to \eta$ & $0.10$ & $0.15$ & $<0.5$ \\
	$D^+\to \eta'$ & $0.016$ & $0.019$ & $<1.1$ \\
	$D_s^+\to K^0$ & $0.20$ & $0.32$ &  \\
\hline 
\end{tabular}
\end{table}
It is very interesting that our model extrapolated with a double
pole gives branching ratios for $D \to P \ell \nu_{\ell}$ in rather good
agreement with experimental results for the already measured decay
rates. It is also obvious that the single pole fit 
gives the rates up to a factor of two larger than the experimental results. 
Only for decays to $\eta$ and $\eta'$ as given in  
Table~\ref{PP_results_table}, an agreement with experiment of the double pole 
version of the model is not better but worse than for the single pole case.
\par
\begin{figure}[!b]
\begin{tabular}{cc}
\scalebox{0.7}{\includegraphics{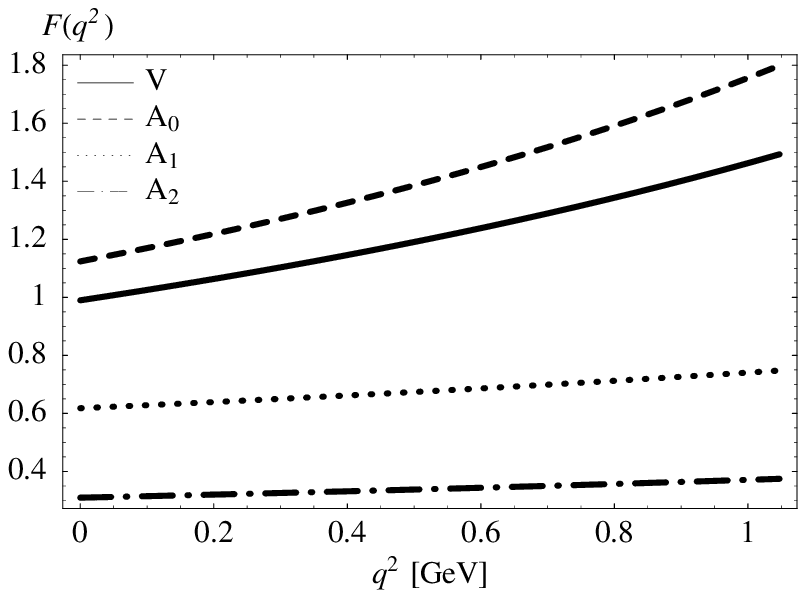}} &
\scalebox{0.7}{\includegraphics{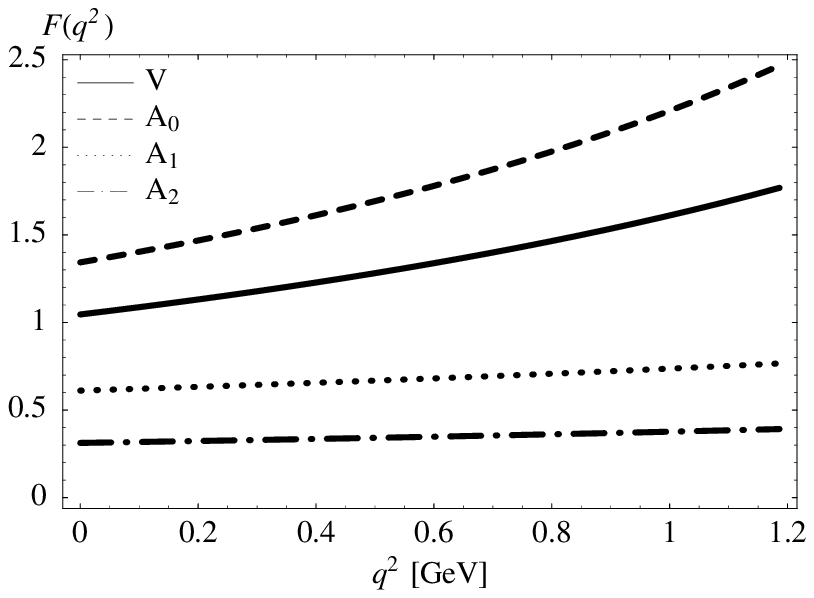}} \\
\multicolumn{2}{c}{\scalebox{0.7}{\includegraphics{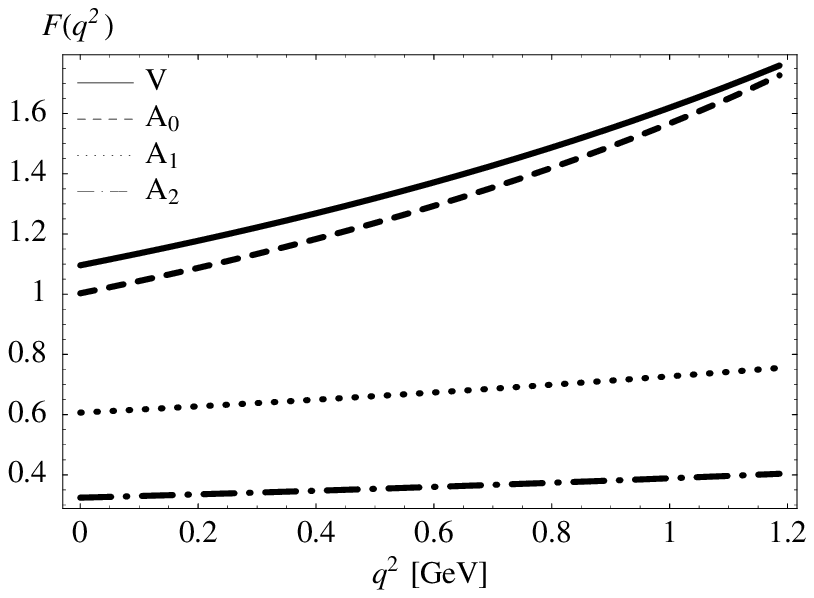}}}
\end{tabular}
\caption{\label{FplotDKs}$q^2$ dependence of the $D^0 \rightarrow K^{*-}$ (upper left), $D^0 \rightarrow \rho^{-}$ (upper right) and $D_s \rightarrow \phi$ (lower) transition form factors.}
\end{figure}

We next draw the $q^2$ dependence of all the form factors for 
the $D^0\to K^{-*}$, $D^0\to \rho^-$ and $D_s \to \phi$ transitions. The results are 
depicted in Fig.~\ref{FplotDKs}.
Our extrapolated results for the shapes of the $D\to V$ semileptonic form factors agree well with existing theoretical studies~\cite{Ball:1991bs, Melikhov:2000yu, Abada:2002ie,   Ball:1993tp}, while currently no experimental determination of the form factors' shapes in these decays exists.  
\par
We complete our study by calculating branching ratios and partial decay width 
ratios also for all relevant $D \to V l \nu_l$ decays. 
They are listed in Table~\ref{table_results} together with known 
experimentally measured values.  
\begin{table}
\caption{\label{table_results} The branching ratios and partial decay width 
ratios for the $D\rightarrow V$ semileptonic decays. Comparison of our model 
fit with experiment as explained in the text.}
\begin{tabular}{l|cccc}
\hline
\tablehead{1}{c}{b}{Decay} & \tablehead{1}{c}{b}{$\mathcal{B}$ (Mod.) [\%]} & \tablehead{1}{c}{b}{$\mathcal{B}$ (Exp.) [\%]} & \tablehead{1}{c}{b}{$\Gamma_L/\Gamma_T$ (Mod.)} &  \tablehead{1}{c}{b}{$\Gamma_+/\Gamma_-$ (Mod.)} \\
\hline 	
	$D_0\to K^*$ & $2.2$ & $2.15 \pm 0.35$~\cite{Eidelman:2004wy} & $1.14$ & $0.22$ \\
	$D_0\to \rho$ & $0.20$ & $0.194\pm 0.039 \pm 0.013$~\cite{Blusk:2005fq} & $1.11$ &  $0.14$ \\
	$D^+\to K_0^*$ & $5.6$ & $5.73\pm 0.35$~\cite{Eidelman:2004wy} & $1.13$ 
	\tablenote{Exp. $1.13\pm0.08$~\cite{Eidelman:2004wy}} 
	& $0.22$ 
	\tablenote{Exp. $0.22\pm0.06$~\cite{Eidelman:2004wy}} 
	\\
	$D^+\to \rho_0$ & $0.25$ & $0.25\pm 0.08$~\cite{Eidelman:2004wy} & $1.11$ & $0.14$ \\
	$D^+\to \omega$ & $0.25$ & $0.17\pm0.06\pm0.01$~\cite{Blusk:2005fq} & $1.10$ & $0.14$\\
	$D_s\to \Phi$ & $2.4$ & $2.0\pm 0.5$~\cite{Eidelman:2004wy} & $1.08$ & $0.21$ \\
	$D_s\to K_0^*$ & $0.22$ &  & $1.03$ & $0.13$ \\
\hline 
\end{tabular}
\end{table}

\par

Finally, we summarize our results: We have investigated semileptonic form factors for $D \to P$ and $D \to V$ decays within an approach which combines heavy meson and chiral symmetry. The form factors are parametrized to satisfy all constraints coming from HQET and SCET. The
contributions of excited charm meson states are included into analysis. The values of form factors at $q_{max}^2$ are calculated using heavy meson chiral Lagrangian.  The second poles of the $F_+(q^2)$, $V(q^2)$ and $A_0(q^2)$ form 
factors are saturated by the presence of the next radial excitations of $D_{(s)}^*$ and $D_{(s)}$.  
The single pole  $q^2$ behavior of the $A_1(q^2)$ form factor is explained 
by the presence of a single $1^+$ state relevant to each decay, while in 
$A_2(q^2)$ in addition to  $1^+$ states one might include their next 
radial excitations. 
\par
The obtained $q^2$ dependence of the form factors is in good agreement with
recent experimental results and existing theoretical studies. The calculated branching ratios are close to the experimental ones. We hope that the ongoing experimental studies will help to shed more light on the shapes of the $D\to P,V$ form factors.

\begin{theacknowledgments}
We thank D. Be\'cirevi\'c for many very fruitful discussions on this subject. S. F. thanks Alexander von Humboldt foundation for financial support and A. J. Buras for his warm hospitality during her stay at the Physik Department, TU M\"unchen, where part of this work has been done. This work is supported in part by the Ministry of Education, Science and Sport of the Republic of Slovenia.
\end{theacknowledgments}

\bibliographystyle{aipproc}   

\bibliography{article}

\end{document}